\documentclass[12pt]{elsarticle}

\usepackage{url}
\usepackage{graphicx}
\usepackage{dcolumn}
\usepackage{bm}
\usepackage{color}
\usepackage{version}
\usepackage{amsfonts}
\usepackage{soul}
\usepackage{float}

\usepackage{mathtools}

\DeclarePairedDelimiter\floor{\lfloor}{\rfloor}

\newcommand{\ba}{\begin{eqnarray}}
\newcommand{\ea}{\end{eqnarray}}
\def\be{\begin{equation}}
\def\ee{\end{equation}}

\excludeversion{details}

\begin{document}

\title{Exploring neural network training strategies to determine phase transitions in frustrated magnetic models}

\author{I. Corte}
\address{Departamento de F\'isica, Universidad Nacional de La Plata,
C.C.\ 67, 1900 La Plata, Argentina.}
\address{Departamento de Ciencias B\'asicas, Facultad de Ingenier\'ia Universidad Nacional de La Plata, La Plata, Argentina}

\author{S. Acevedo}
\address{IFLP - CONICET, Departamento de F\'isica, Universidad Nacional de La Plata,
C.C.\ 67, 1900 La Plata, Argentina.}
\address{Departamento de Ciencias B\'asicas, Facultad de Ingenier\'ia Universidad Nacional de La Plata, La Plata, Argentina}

\author{M. Arlego}
\address{IFLP - CONICET, Departamento de F\'isica, Universidad Nacional de La Plata,
C.C.\ 67, 1900 La Plata, Argentina.}

\author{ C. A.\ Lamas}
\address{IFLP - CONICET, Departamento de F\'isica, Universidad Nacional de La Plata,
C.C.\ 67, 1900 La Plata, Argentina.}
\address{Departamento de Ciencias B\'asicas, Facultad de Ingenier\'ia Universidad Nacional de La Plata, La Plata, Argentina}




\begin{abstract}
The transfer learning of a neural network is one of its most outstanding aspects and has given supervised learning with neural networks a prominent place
in data science. Here we explore this feature in the context of strongly interacting many-body systems.
Through case studies, we  test the potential of this deep learning technique to detect phases and their transitions in
frustrated spin systems, using fully-connected and convolutional neural networks. In addition, we explore a recently-introduced technique, which is at the
middle point of supervised and unsupervised learning. It consists in evaluating the performance of a neural network that has been deliberately ``confused"
during its training. To properly demonstrate the capability of the ``confusion" and transfer learning techniques, we apply them to a paradigmatic model of
frustrated magnetism in two dimensions, to determine its phase diagram and compare it with high-performance Monte Carlo simulations.

\end{abstract}
\maketitle

\section{Introduction}

The field of machine learning, in particular deep learning, has gained a prominent place in practically all areas associated with technology
\cite{lecun2015deep}. This is the result of the symbiosis between data generation, computing power, and algorithm development.
What makes machine learning techniques especially useful in many applications is the automatic search for patterns and underlying models in the data.
These models are then used to classify, predict, generate, and make decisions about new events or data.

The characteristic elements of the many-body interacting systems, such as high dimension, correlations, symmetries, and phase transitions, naturally emerge
in data science and machine learning \cite{DeepLearningBible,hastie2009elements,language-power-law-corr,Phase-Transitions-in-ML-Book}.

For this reason, it is evident that neural network techniques, which have been fundamental in data science and machine learning,
will also play an important role in the physics of interacting many-body systems. This fact is reflected in some recently articles \cite{MEHTA20191,carleo2019machine,Carrasquilla-Review}.
Our work points in this direction, taking a step forward in the implementation of neural networks to study frustrated magnetism {(systems where cannot simultaneously minimize the energy contribution of all the magnetic couplings).}

In this paper, motivated by the work of Carrasquilla and Melko \cite{carrasquilla2017machine} and the research carried out recently
\cite{wang2017machine, wang2018machine, beach2018machine,ch2017machine,zhang2017machine,Potts2020,ponte2017kernel,suchsland2018parameter,greitemann2019probing,
ch2018unsupervised,broecker2017quantum,huembeli2018identifying,liu2019learning,hsu2018machine,broecker2017machine,
zhang2019interpretable,ni2019machine,rem2019identifying,wetzel2020discovering,wang2016discovering}, to cite a few, we analyze a variety of correlated classical spin models using neural networks.
This provides a complement to more traditional methods, which include diverse analytical and computational
tools \cite{Auerbach:1994,Book-Computational-Many-Particle-Physics}.

\begin{figure*}[t!]
\begin{center}
\includegraphics[width=0.9\textwidth]{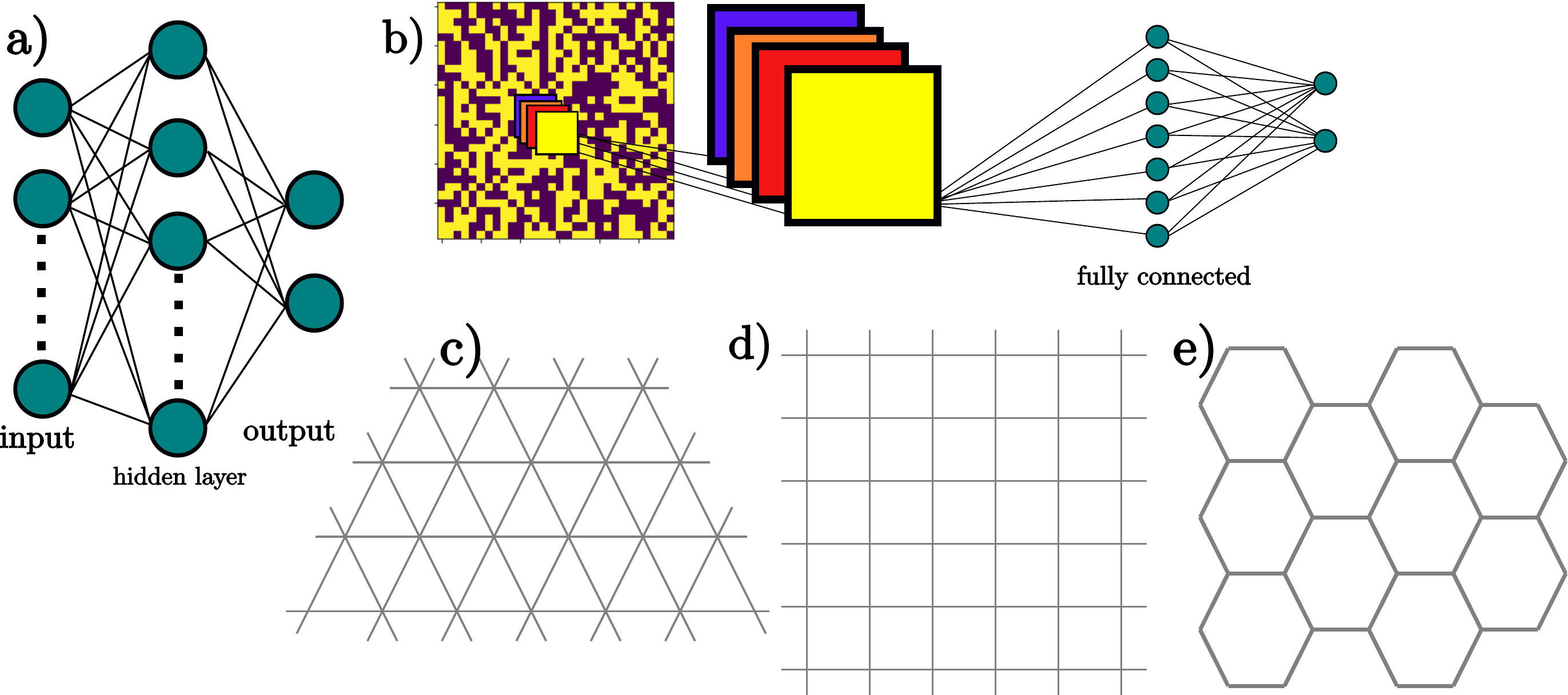}
\end{center}
\caption{\label{fig:CNN}
Schematic representation of the neural networks and lattices implemented in this work. a) Fully-Connected Neural Network, or Dense Neural Network (DNN). b)
Convolutional Neural Network (CNN). The input features correspond to 2D system configurations which are re-shaped into a 1D structure for the DNN,
while in the CNN the input is the 2D ``image" of the configuration. In both cases, the two-neurons output layer corresponds to order-disorder probabilities. c) Triangular, d) Square and e) Honeycomb lattice.}
\end{figure*}

This work focuses on the flexibility and the generalization power provided by the use of
fully-connected neural networks, also called dense neural networks (DNN)\cite{chollet2017-Book} and convolutional (CNN) neural networks.
These architectures are schematically indicated in Fig. \ref{fig:CNN} (a) and (b), respectively.
We aim to explore the possibilities that these techniques can offer in the study of correlated systems, beyond a particular and precise determination
of a phase transition in a specific model and lattice.

{
A key ingredient in this study is the incorporation of frustration into the system.
That is why we choose to study purely antiferromagnetic systems where frustration emerges naturally.
Unlike previous studies where the performance of neural networks in ferromagnetic systems is explored,
when studying a frustrated system, whose order parameter is more complex, some architectures begin to fail and more complex neural networks must be used.
}

Throughout this paper, different types of spin lattices are explored (Fig. \ref{fig:CNN} (c-e)), together with several groups of
hyperparameters \cite{DeepLearningBible}, to test the performance of a given neural network against a classification task.
The data in our work is synthetically generated by Monte Carlo simulations \cite{Landau-Binder-MC-Book}, which we implement for the different models
and lattices studied.

For the supervised-network training, we test different types of input data, such as local spin configurations and correlations.
This allows us to analyze the advantages and possibilities offered by the different types of features and their relevance in each particular case.

Most of the computations performed in this work are open source \cite{chollet2015keras}. The corresponding codes are freely accessible via GitHub and data
is available upon request to the authors. We have relegated most of the technical details about training and network architectures to the Appendix.

A significant part of this work is devoted to studying the ``transfer learning'' ability of neural networks in this kind of phase classification.
Transfer learning consists of exploiting the predictive capacity of a neural network beyond the context
in which it has been trained.
In the framework of machine learning, the concept of transfer learning is broad \cite{DeepLearningBible}.
It may be \textit{indirect}, such as fine-tuning transfer learning, in which a pre-trained model in a certain domain is then fine-tuned in another
domain, and feature-based transfer learning, in which certain intermediate features of a model are transferred to another domain.
{ In order to clearly differentiate the scenarios where we will test the ability to generalize the knowledge obtained in the training
for a neural network, we will use different labels. } We use the label \textit{direct} transfer learning, where the model trained in one domain
is directly applied to another domain.
This presents several levels of complexity. At a more basic level, networks trained in a restricted sector of a given model and lattice,
for example at low and high temperatures, are used to predict its transition temperature \cite{broecker2017machine}.
At a higher level, networks trained with a certain model and lattice are used to predict properties, such as critical transition temperature,
of the same lattice but in another model\cite{ch2017machine}, {  we label this scenario as \textit{model} transfer learning }.
Further increasing the complexity, networks trained with a given model and lattice are used to predict the same properties in a different model and
lattice \cite{carrasquilla2017machine}, {  we label this scenario as \textit{model-lattice} transfer learning }.
{ We note that, although these are not standard names, their use will make the explanation of the results clearer.}

{As we have mentioned, the different transfer processes has been used in other works. However, the novelty here is its application to more complex systems, where frustration induces high degeneracy, representing a greater challenge to the power of generalization of the network as a classifier. }
We analyze these transfer learning processes on different realizations of the antiferromagnetic Ising model, including first- and second-neighbor interactions on the square, honeycomb, and triangular lattices.

The last part of our work explores a technique that could be considered a middle point between supervised and unsupervised learning.
This method\cite{confusion} of ``learning by confusion'' exploits the variability of a network performance that has been deliberately trained with incorrect
labels.
The advantage of this technique is that it does not need the correct labels for learning and can detect phase transitions, or at least significant pattern
variations that suggest such a transition.
This is the reason why it can be considered an unsupervised learning method. With the confusion technique, we teach a network to classify the phases of the
AFM Ising model on the square lattice with second-neighbors interactions, which is an archetype of classical frustrated two-dimensional magnetism.
This study complements the phase diagram determined with a CNN by \textit{model transfer}.

For completeness, in the following we present a brief discussion of the prominent aspects of the Ising model family of interacting systems that we are going
to analyze using neural networks.

{We consider the Ising model in the absence of magnetic field described by the following Hamiltonian $H$, representing the energy of the system}
\begin{equation}
  \label{eq:Ising-Hamiltonian}
 H=\sum_{ i,j }J_{i,j} \sigma_i\sigma_{j},
\end{equation}
in several two-dimensional lattices, where $J_{i,j}$ is the coupling between spins  {  $\sigma_i$ and  $\sigma_j$ (which can take values $\{\pm 1\}$) } on sites $i$ and $j$.
{ Throughout the text, we simplify the notation by denoting first neighbors couplings as $J$  and we use $J_2$ for second neighbors.}
Here, we will be particularly focused on the AFM case ($J_{i,j}>0$), due to its higher complexity and richness.  {The ferromagnetic case ($J_{i,j}<0$), although equally important, offers fewer difficulties and has been much more explored using neural networks, driven by Carrasquilla and Melko's work \cite{carrasquilla2017machine}, among others.}
The two-dimensional Ising model on the square lattice with first-neighbor interactions
$J$ was analytically solved by Onsager in 1944 \cite{onsager1944crystal}.
He showed that there is an order-disorder phase transition for the infinite square lattice at the critical temperature
$T_{c}=\frac{2|J|}{\ln(1+\sqrt{2})} \simeq 2.269 |J|$ in units of Boltzmann constant $k_B$. On the honeycomb lattice, the critical
temperature is also analytically available \cite{Ising-Exact-Honeycomb}, $T_{c}=\frac{2|J|}{\ln(2+\sqrt{3})} \simeq 1.519 |J|$.
In both cases the result is valid for FM $(J<0)$ and AFM $(J>0)$ interactions.
Solutions for general lattice topologies and couplings are obtained by means of series expansions and numerical methods,
such as Monte Carlo \cite{Landau-Binder-MC-Book}.

%

The concept of frustration, which accounts for the impossibility of simultaneously minimizing all the couplings in the Hamiltonian
from Eq. \ref{eq:Ising-Hamiltonian}, takes the Ising model to a higher order of complexity.
Frustration can take place in different ways, either due to the structure of the lattice itself or due to the inclusion of interactions beyond first
neighbors, with additional finite couplings $J_2, \, J_3$, etc.
 {The AFM Ising model with first-neighbor couplings on the triangular lattice is an example of the first type. In this case, it is not possible to arrange the spins so that all interactions between them are antiparallel. Even in a single triangle of the lattice, if two antiparallel spins are placed on two vertices, the third vertex cannot be antiparallel to the other two. The system is ``frustrated'' because it cannot simultaneously minimize the energy contribution of all three couplings. As a consequence, there is no single way to minimize the system energy and it becomes degenerate at $T = 0$ \cite{wannier1950-Ising-AFM-Triangular}. This is a crucial difference with the FM counterpart, where the energy is minimized by placing all the spins in parallel. }\\
On the other hand, the AFM Ising model with second-neighbor couplings on the square and honeycomb lattice illustrates the second case of frustration.  {Here the first neighbors can be arranged antiparallel, but if second neighbors are included this cannot be satisfied anymore.} These are some of the cases we analyze in this work.

A direct consequence of frustration is the high degeneracy of the ground state, resulting in a wider variety and complexity of structures \cite{Frustrated-Magnetism-book}.
This makes frustrated systems ideal to explore the power of classification and generalization of neural networks in condensed matter interacting systems.
%


%

\section {Training I: Local transfer}

In this section, we begin our study of the phase diagrams with neural networks by exploring their performance to classify ordered and disordered phases
by \textit{local transfer}.
We also present the general procedure for generating and labeling data, as well as the training-test scheme that we follow in the supervised learning
part of the work.

\subsection {Honeycomb lattice}

To begin with, we considered the AFM Ising model on the honeycomb lattice. We start by considering the simplest neural network, a DNN.

We found that to classify the ordered and disordered phases in the AFM Ising model on the honeycomb lattice, it is enough to use a DNN with a single hidden layer of 16 neurons. As input variables, in this case, we use the local spin configurations that we compute from Monte Carlo simulations \cite{Landau-Binder-MC-Book}. Since this is a binary classification problem involving ordered and disordered configurations,
we employ the binary cross-entropy cost function, together with an $ L_ {2} $ regularization to further control overfitting  {(see Appendix for further details).}

Monte Carlo generation of tagged data is performed as follows. We run 400 independent simulations starting from the high-temperature phase.
The whole temperature range is partitioned, and for each temperature, the spin configuration and the temperature are saved once equilibrium is reached.
Data with $T<T_c$ is labeled with 0 and data with $T>T_c$ is labeled with 1.

The data generated in the simulations is split $70\%$ for training and validation and $30\%$ for the test (prediction).
Note that the temperature information is not introduced explicitly during training, since the DNN only uses the local spin configurations as input features and the labels 0 and 1 for minimizing the cost function. The temperature is used only in the test stage to analyze the performance of the classification and prediction of the critical temperature.

\begin{figure}[t!]
\begin{center}
\includegraphics[width=0.9\textwidth]{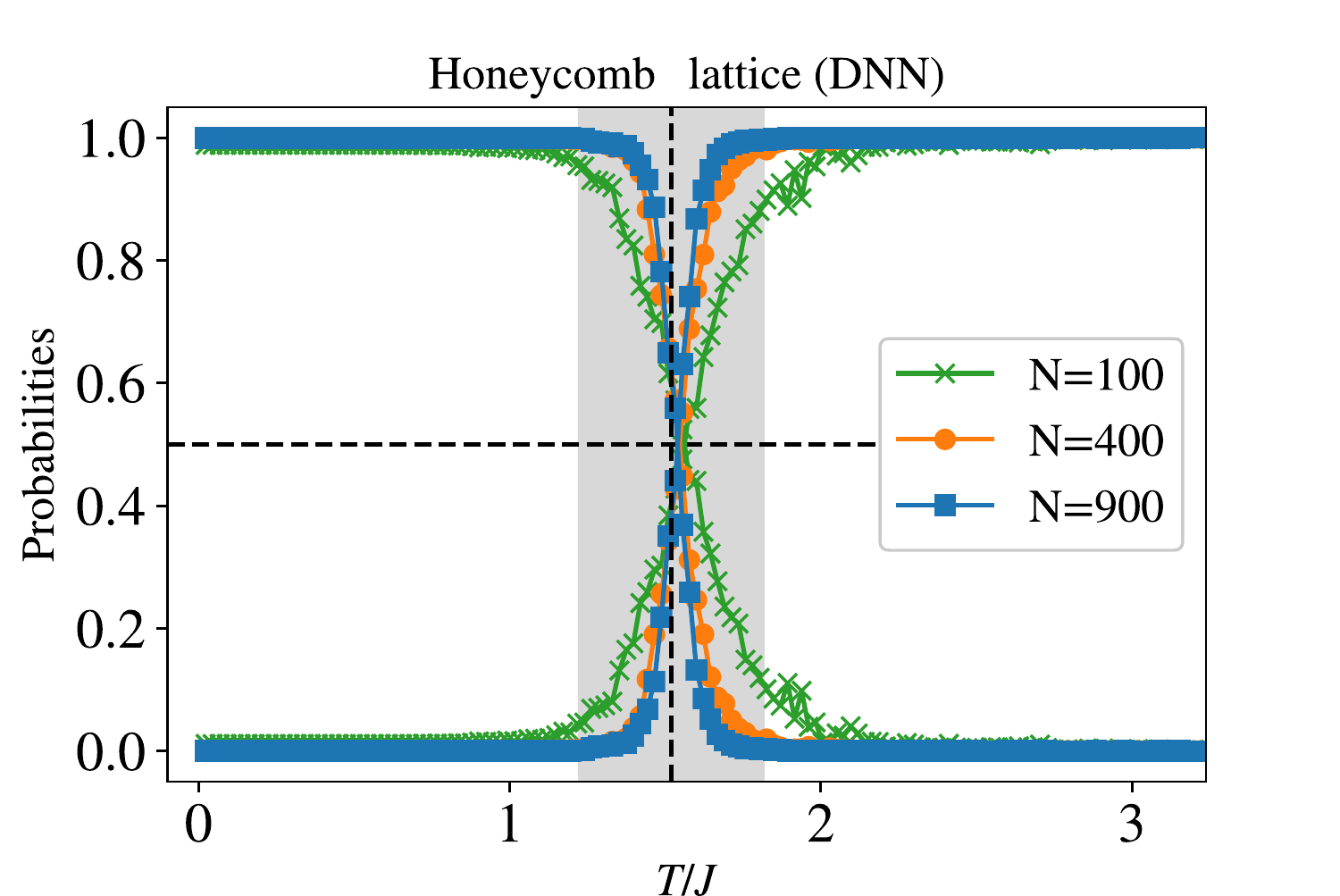}
\end{center}
\caption{\label{fig:AFM-Hon-NN-Ventana}Local (non-frustrated) DNN transfer. DNN output layer probabilities averaged over the test dataset as a function of $T$ for the AFM-Honeycomb lattice.
The lattice sizes are $N=100,\,400$ and $900$ sites. The training data corresponds to the ranges $0.02 < T/J < 1.22 $ and $1.82 < T/J < 4.53$. The shaded gray region represents the hidden set of data during training, and the vertical dotted black line shows the analytical result for the honeycomb lattice critical temperature in the thermodynamic limit.
Validation accuracy, i.e. the ratio of correct predictions over the total number of predictions at the validation stage is 0.99, for $N=100,\,400$ and $900$ sites (see Appendix for further details).}
\end{figure}

 {To be able to say if a prediction is correct or incorrect using supervised learning architectures, one has to compare it with the sample label. There are several ways to measure the performance of a neural network. Here we employ the \emph{accuracy}, which is defined as the ratio between the number of well-classified samples and the total number of samples. The accuracy may depend on several factors such as the characteristics of the system near the transition temperature, the size of the training set, the geometry of the network used, etc. In any case, a validation accuracy of 0.99, for instance, tells us that the DNN correctly classified 99 out of 100 snapshots of the validation set (which is not used for training).}

To test the network's ability to predict beyond the data in which it is trained, we have trained the network in a range of temperatures that excludes a
window of width $w$, centered at the transition temperature $T_{c}$. This case illustrates the \textit{local transfer} process.
{ Dense neural networks have shown that they can determine the transition temperature of a ferromagnetic system in this way. They show that they can learn simple order
parameters such as magnetization. However, in the antiferromagnetic case, the order parameter is not so simple and depends on the lattice geometry. }

Fig. \ref{fig:AFM-Hon-NN-Ventana} presents the results obtained for the classification of the ordered and disordered phases in the AFM Ising honeycomb
lattice, for three different lattice sizes N of $100$, $400$, and $900$ spins.
In particular, it shows the probability of belonging to each phase as a function of temperature.
This probability is obtained for each temperature by averaging the values predicted by the network on test data (not used for training) corresponding to
this temperature. The shaded gray region indicates the range
of data excluded for training (\textit{local transfer}).
Since the classification is binary, only the probability curve of one class is required for each size. For example, the blue curve starting at the top
left of Fig. \ref{fig:AFM-Hon-NN-Ventana} indicates the probability of belonging to the ordered class for $N = 900$, while the lower blue curve to the
left indicates the probability of belonging to the disordered class. Both curves add probability 1 and therefore are not independent (they are symmetric with respect to the line $p = 0.5$). However, throughout work, we show both probability curves as a visual guide to locate the predicted critical temperature,
i.e. the crossing between both curves. The analytical critical temperature for the Ising honeycomb lattice (in units of $k_B$), $T_{c} \simeq 1.519 J$,
is indicated by the black dotted line in Fig. \ref{fig:AFM-Hon-NN-Ventana}.

First, note that the DNN clearly separates the ordered and disordered classes for temperatures far from the transition This is reflected in a probability
prediction of approximately 1 (0) for the ordered (disordered) phase on the left and vice versa on the right.

However, as the temperature approaches $T_{c}$, the DNN finds it more difficult to differentiate between the two phases, which is indicated by the approach
of the curves to the intermediate zone of the Fig. \ref{fig:AFM-Hon-NN-Ventana}.
The crossing of both curves at $p = 1/2$ \textit{defines} the transition temperature predicted by the DNN for each size.

As it can be observed in Fig. \ref{fig:AFM-Hon-NN-Ventana}, the prediction of the critical temperature is very close and slightly shifted to the right of
the critical value, for each size.
However, for $N = 100$ (green curves) the loss of predictive power falls faster and is noisier than in the other sizes.
This reflects the limitations and finite-size effects when using a smaller lattice.

The most important aspect to emphasize here is the performance of the DNN in the intermediate zone around the critical temperature, denoted by the gray area.
Given that this intermediate zone is deliberately removed from training, the predictions made in this region manifest the generalizing power of the network beyond training. In other words, the DNN predicts the critical temperature with high accuracy despite never having seen data from the transition zone.
This example illustrates the efficiency of \textit{local transfer} with DNNs on non-frustrated lattices.

\subsection {Triangular lattice}

To test the performance of DNNs in a more complex context, we analyzed the AFM Ising model on the triangular lattice.

In the thermodynamic limit, this system has finite zero-point entropy, is disordered at all temperatures, and has no Curie
point \cite{wannier1950-Ising-AFM-Triangular}.
However, the Monte Carlo simulation of the  model shows a maximum in the specific heat at a certain temperature $ T^* $, below which short-range
correlations emerge. The task of the neural network, in this case, is to differentiate low and high-temperature configurations, i.e. on both sides of $T^*$,
when trained with samples of Monte Carlo simulations, excluding a window around $ T^* $.

\begin{figure}[t!]
\begin{center}
\includegraphics[width=0.9\textwidth]{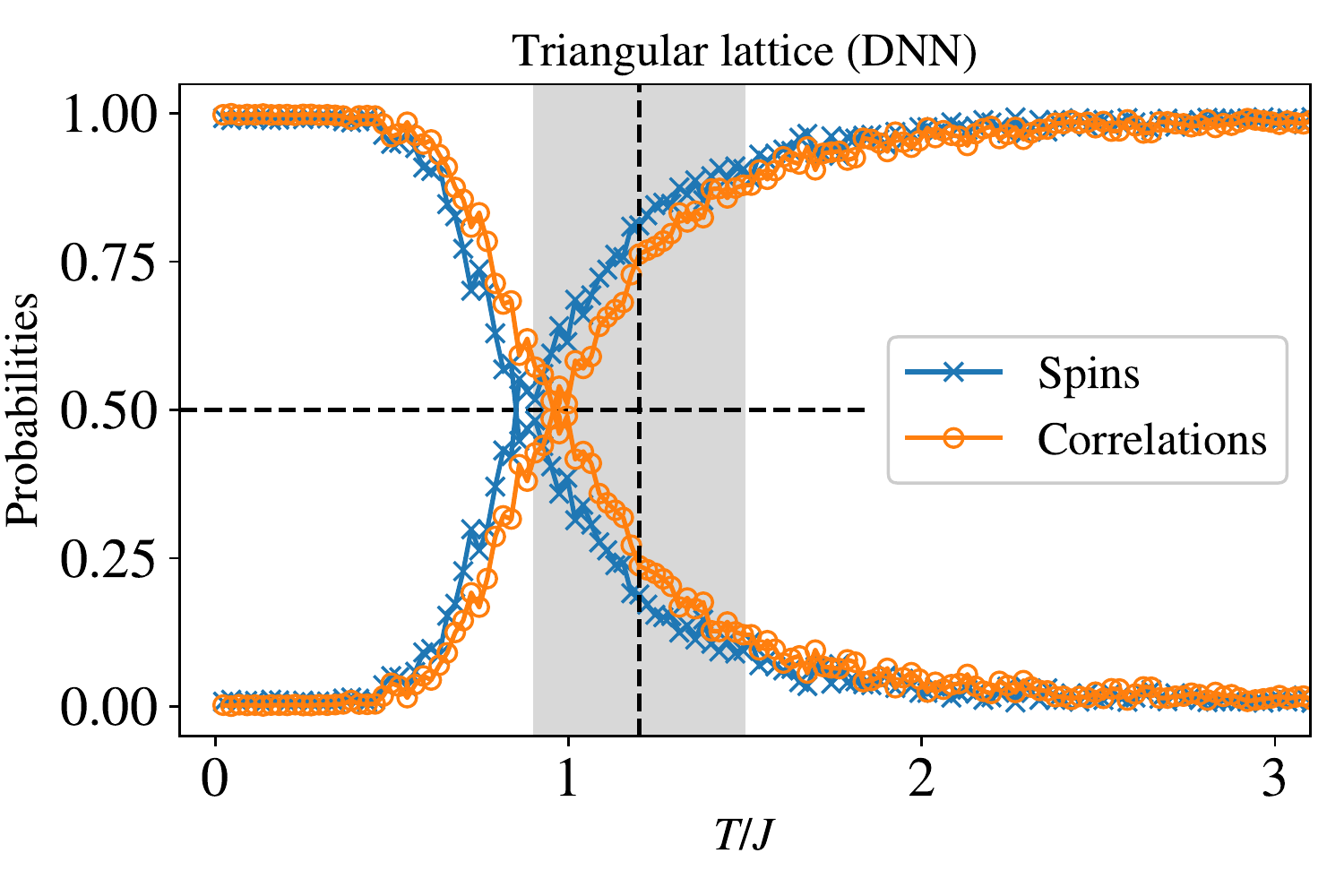}
\end{center}
\caption{\label{fig:AFM-triang-NN-Ventana}Local (frustrated) DNN transfer. DNN output layer probabilities averaged over the test set, as a function of temperature for the AFM-Triangular lattice.
The blue crosses and orange circles correspond to training with local spin configurations and correlations, respectively.
The shaded gray region indicates the range of data that was excluded for training.
The vertical dashed black line corresponds to the temperature $T^*$ where the specific heat of the system reaches its maximum value in the Monte Carlo simulation.
In both cases the DNN
underestimates $T^*$, i.e. the crossing of both probability curves due to the additional complexity that frustration introduces in the triangular lattice. Validation accuracy: 0.97 for spin configurations and 0.98 for correlations.}
\end{figure}

In the following, we set system sizes to $N=900$ sites.
Fig. \ref{fig:AFM-triang-NN-Ventana} shows our results for the triangular lattice, trained with two different input features, using a DNN with a hidden
layer of 32 neurons.
The blue line shows the results obtained by training the network with local spin configurations, while the orange line corresponds to the ones obtained by
training with correlations.
The latter has been shown to be useful in multicomponent models, such as the Potts model \cite{Potts2020}. For the Ising model we have evaluated the
correlations $C_{x,y} = \sigma_{x,y} \sigma_{L/2,L/2},$ relative to the center $(L/2, L/2)$ of a $L\times L$ lattice with periodic boundary conditions.
Subscripts $x, y$  denote the spin position, relative to the vectors of the lattice unit cell. As can be observed in Fig. \ref{fig:AFM-triang-NN-Ventana},
in both cases, the DNN underestimate the position of  $T^*$, indicated by the dashed vertical line in the figure.
Using local spins as input features, the DNN predicts the $ T^* $  that falls slightly outside the left zone excluded from training (shaded gray area).
The use of correlations in the training process improves the prediction of $ T^* $, which falls inside the zone excluded from training. However, the additional complexity that frustration introduces in the triangular lattice reduces the DNN's performance to determine $ T^* $ in both cases. The overall performance of the network is quite insensitive to differences in the choice of hyperparameters of the network or alternative ways of evaluating correlations \cite{Potts2020}. The AFM Ising model on the triangular lattice shows the limitations of DNNs
in the presence of frustration and suggests the usage of higher-complexity neural network architecture would be more appropriate for this problem.

To address the previously-mentioned problem of limited DNN's performance in a frustrated lattice, we analyzed the Ising AFM model on the triangular lattice
with a CNN.
Convolutional networks are well-adapted to classify images, which in our case are the local spin configurations for each temperature value.
The CNN extracts the relevant features from the input image through a successive application of preprocessing filters and then these feature maps serve as input for a final dense network. The filters and DNN parameters are learned during training.

\begin{figure}[t!]
\begin{center}
\includegraphics[width=0.9\textwidth]{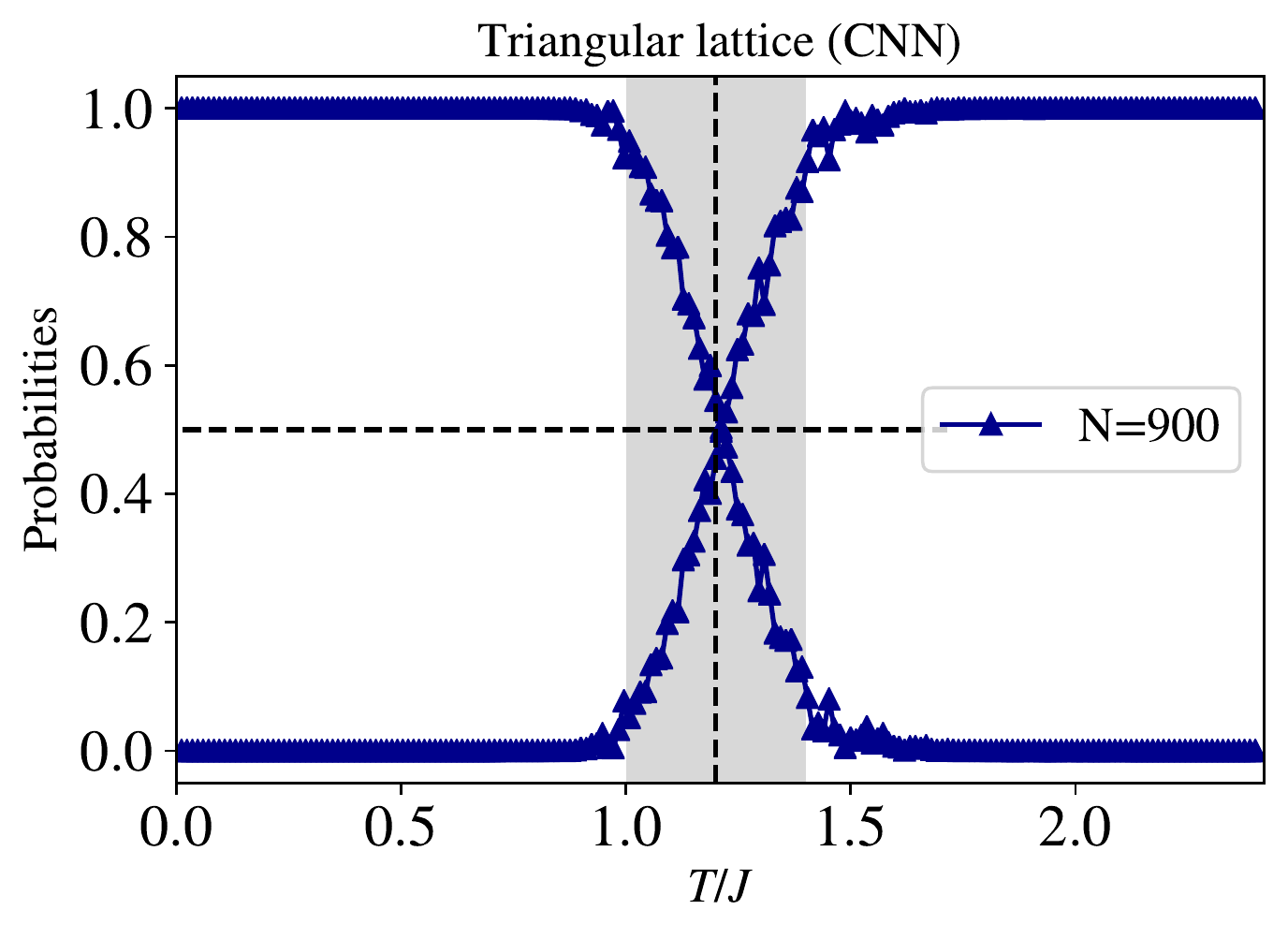}
\end{center}
\caption{\label{fig:AFM-traing-CNN-Ventana}
Local (frustrated) CNN transfer: Output probabilities as a function of temperature, for the AFM-triangular lattice using a CNN.
The shaded gray area corresponds to the range of data that was not used for training.
The vertical line denotes the value of $T^*$, where the specific heat of the system reaches its maximum value with $N=900$ sites.
Validation accuracy of 0.99.
}
\end{figure}

Fig. \ref{fig:AFM-traing-CNN-Ventana} shows the results obtained using a CNN for the classification of low and high-temperature configurations of the
AFM Ising model on the triangular lattice.
The results were obtained by implementing a CNN with a convolutional layer of 10 filters, followed by a dense layer of 10 neurons.
As before, we have applied the strategy of training excluding a window around $T^*$, limited by the shaded gray area. Unlike the DNN results in
Fig. \ref{fig:AFM-triang-NN-Ventana}, the probability curves cross around $T^*$, indicated by the dashed vertical line.

Since the CNNs show higher performance than DNNs in frustrated systems, in the following sections we will concentrate on the implementation of CNNs in
these systems. In particular, we will analyze other aspects associated with the performance of CNNs and their power of generalization, not only at
the local transfer level as we have done here. We will fully explore the potential of CNNs to predict transitions in other models as well as model and lattices in which the network has not been trained, i.e. CNN \textit{model transfer}, and \textit{model-and-lattice transfer} respectively.

 {We would like to end this section by making a general comment on the number of layers and the depth of the networks used.
It is possible to increase the number of layers in the network, however it does not significantly improve the accuracy for the cases under study. In contrast, as the neural network gets deeper the number of trainable parameters augments exponentially, which significantly increases the computational cost and the likelihood of overfitting.
In our cases, the neural network must be able to detect ordered phases (or low temperature regimes) outside the training range. For this reason, it is very important to avoid overfitting. 
}

\section{Training II: Model transfer}

As we discussed in the previous section, the CNNs can be very powerful to differentiate between low and high-temperature configurations in frustrated
models as the AFM-triangular lattice, by training at temperatures away from $T^*$ (local transfer).
One step further would be to apply pre-trained neural networks in a given model and lattice to predict transitions in other models or lattices.
In this section, we address \textit{model transfer} by analyzing the performance of a pre-trained CNN to study another model, different from the one
in which it was trained.

\begin{figure}[t!]
\begin{center}
\includegraphics[width=0.9\textwidth]{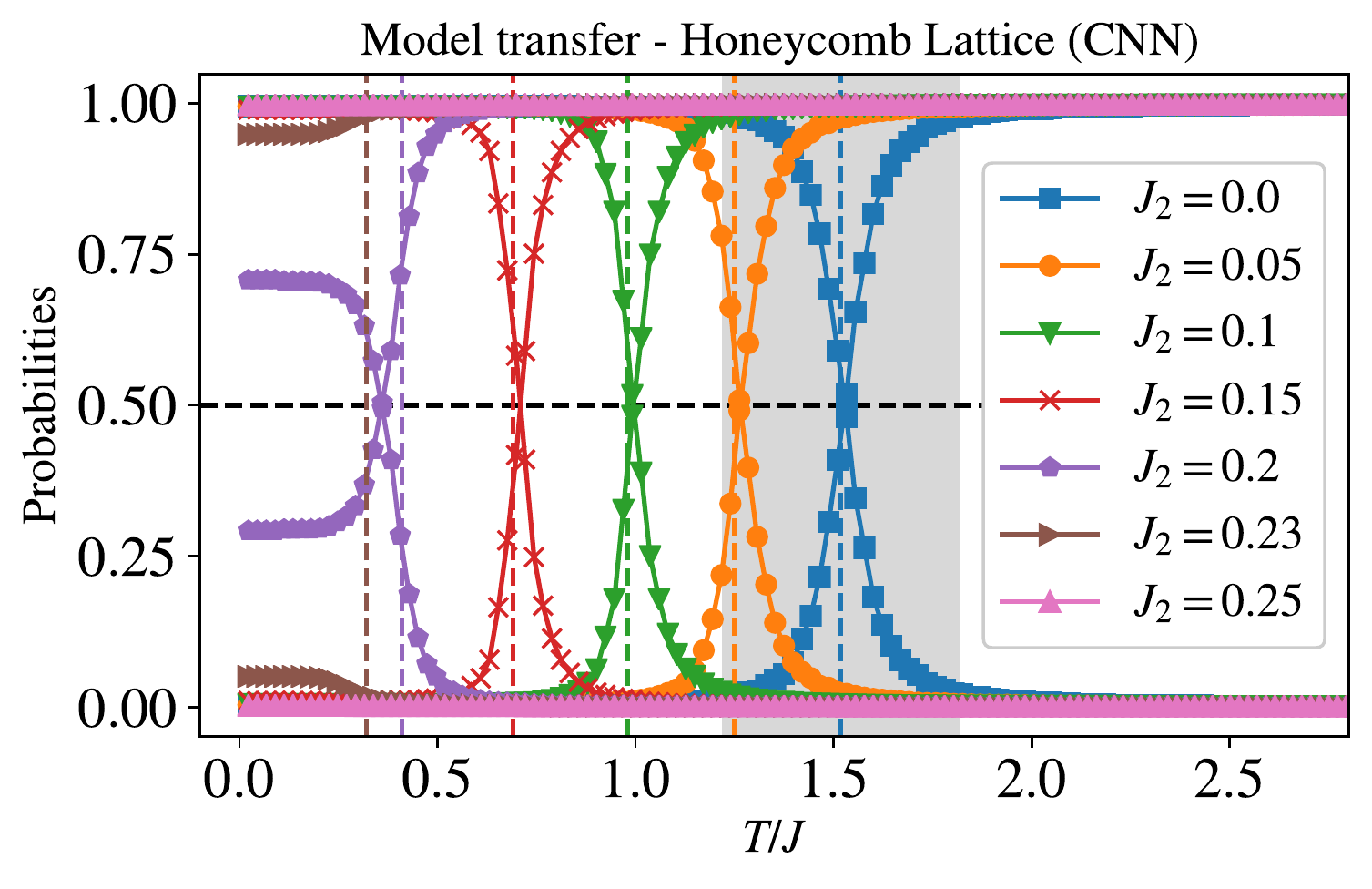}
\end{center}
\caption{\label{fig:model_tranfer_honeycombe_J2_Neel}
Model transfer on the honeycomb lattice: CNN-predicted probabilities for ordered and disordered phases as a function of temperature in the AFM-honeycomb lattice.
The CNN was trained using data for which $J_{2}/J=0$ and temperatures outside the shaded gray area. Vertical dotted lines represent the Monte Carlo estimation of the transition temperatures.
The network is able to correctly estimate the transition temperature of the model with second-neighbor interactions, despite being trained in the nearest-neighbor model. As $J_2/J$ approaches $0.25$ the change in the probability output suggests the presence of a phase transition. In the legend the $J_2$ values are in units of $J$. Validation accuracy: 0.99.
}
\end{figure}

We start by training the CNN on the AFM Ising model on the honeycomb lattice with first-neighbor couplings ($J$), excluding a region around the critical
temperature. To this end, we map the honeycomb lattice to a square array as detailed in Appendix \ref{sec:mapping}. This allows us to carry out comparisons
and the lattice transfer in this work.
Next, we evaluate its performance on data generated in the honeycomb lattice with first- and second-neighbor interactions, i.e. finite $J_2$.
If the ratio of second- to first-neighbor coupling $J_{2}/J$ is small, the system remains in the same phase and it is reasonable to think that the CNN
will be able to recognize the order. However, the transition temperature is a function of $J_{2}/J$ and a priori it is not clear that the CNN can
correctly detect the $ T_c $ behavior.

In Fig. \ref{fig:model_tranfer_honeycombe_J2_Neel} we show results of the probabilities predicted for the N\'eel ordered and disordered phases as a
function of temperature. Each curve corresponds to a different value of the frustrating coupling $J_{2}/J$ and the dashed vertical lines indicate the
respective Monte Carlo estimations of the critical temperatures (See Appendix for details).

Let us recall that the CNN has not been trained with the presence of frustrating interactions, nor near the transition of the first-neighbor model,
and yet for small values of $J_{2}$ the convolutional network prediction is remarkable.

The system has a low-temperature phase transition at $J_2/J=0.25$, from the N\'eel phase to another phase where there is no long-range order due to the
ground state degeneracy \cite{Ising-Honeycomb-frustrated}.
As $J_2/J$ approaches 0.25 in Fig. \ref{fig:model_tranfer_honeycombe_J2_Neel} the CNN probability output changes progressively.
At $J_2/J=0.25$, there is zero probability prediction that the system is ordered for all temperatures.
As the neural network was trained to learn the difference between N\'eel order and disorder, this outcome suggests that the system abandons the N\'eel
ordered phase at $J_2/J=0.25$.

\begin{figure*}[t!]
\begin{center}
\includegraphics[width=0.9\textwidth]{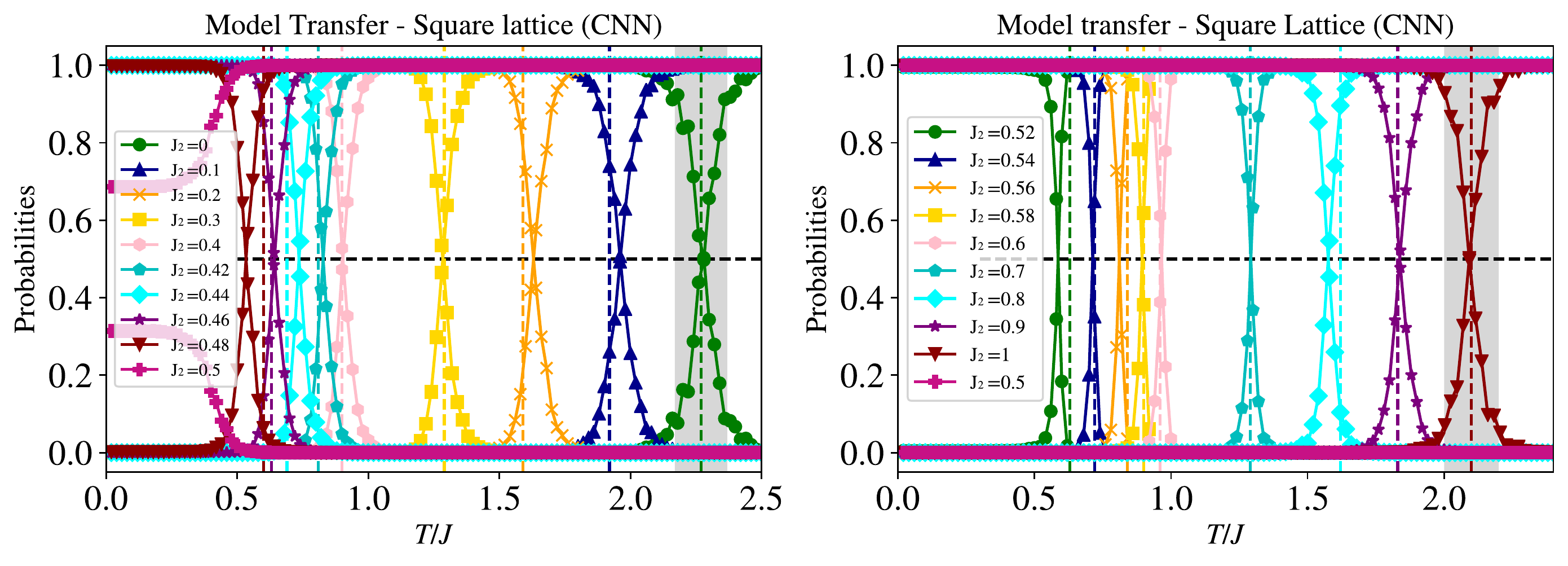}
\end{center}
\caption{\label{fig:transfer-square-juntas}
Model transfer on the square lattice:  Left: CNN output layer probabilities for N\'eel ordered and disordered phases for different values of $0 \leq J_{2}/J < 0.5$. The CNN was trained only with $J_2 /J = 0$ data and temperatures outside the shaded gray area. Validation accuracy: 0.98. The CNN captures correctly the behavior of the transition temperature with respect to the parameter that regulates the frustration of the system, $J_{2}/J$.
Right: CNN output layer probabilities for collinear-ordered and disordered phases for different values of $0.5 <J_{2}/J \leq 1$.
The CNN was trained only for $J_2 /J = 1$ and temperatures outside the shaded gray area. Validation accuracy:  0.99. We observe a similar behavior as in the left panel but as the ratio $J_2/J$ decreases. In both panels, for $J_2 \approx 0.5 J$ the change in the probability predictions suggests a phase transition, consistent with the zero-temperature  N\'eel - collinear transition at $J_2/J=0.5$. Dashed vertical lines indicate the Monte Carlo estimation for the transition temperatures. In the legends the $J_2$ values are in units of $J$.}
\end{figure*}

To further explore \textit{model transfer}, we have also analyzed the Ising model with first and second-neighbor couplings on the square lattice,
starting from two different limiting cases.
On the one hand, we trained a CNN with AFM Ising model data on the square lattice with first-neighbor couplings excluding as before the zone around
the transition.
Next we analyzed the network performance on the square lattice, including frustrating second-neighbor interactions $J_2$.
The left panel of Fig. \ref{fig:transfer-square-juntas} shows the results of the network predictions for the transition temperature between
the disordered and N\'eel phases for different values of frustration $J_{2}/J$, whereas the dashed vertical lines indicate the Monte Carlo estimated
critical temperatures.
As can be observed, the CNN predicts accurately the change in the critical temperature for small $J_2/J$. For $J_2/J \approx 0.5 $ the prediction becomes
less accurate.
This is consistent with the $T=0$ transition from N\'eel to collinear phase at $J_2/J=0.5$, characterized by a highly degenerate
ground state \cite{Andreas-Frustrated-Ising-Square-1}.

\begin{figure*}[t!]
\begin{center}
\includegraphics[width=0.9\textwidth]{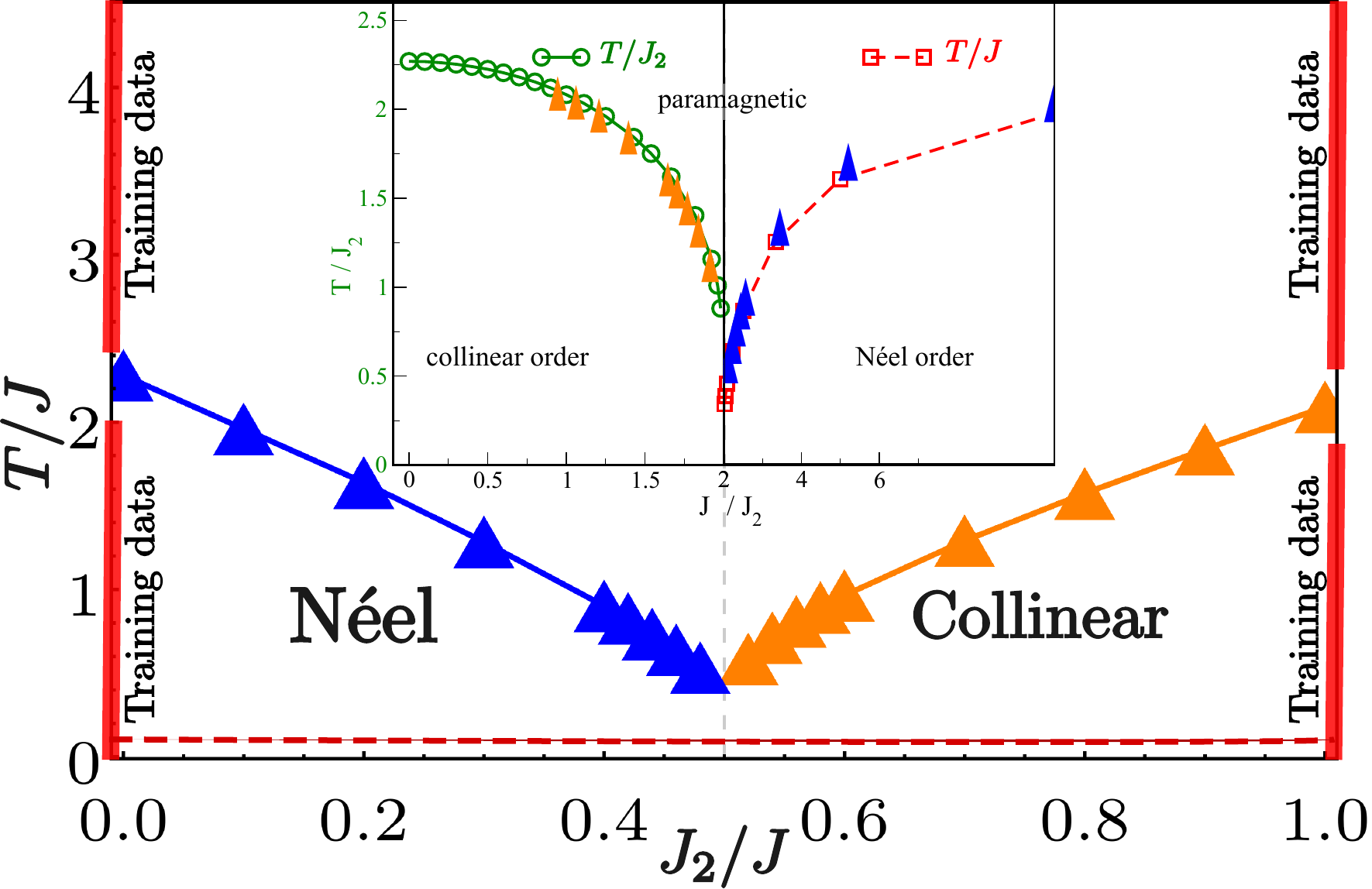}
\end{center}
\caption{\label{fig:pd-inset-andreas}
Frustrated phase diagram by transfer learning. Main panel: CNN predicted phase diagram of the frustrated AFM Ising model on the square lattice with critical temperatures corresponding to the crossings at $p=0.5$ of Fig. \ref{fig:transfer-square-juntas}.
Blue (orange) triangles indicate transition points from the N\'eel (collinear) to disordered phase,
predicted via transfer learning from $J_2/J=0$ ($J_2/J=1$). The horizontal red dashed line indicates the temperature value used for confusion method training shown in Section \ref{sec:confusion}.
Inset: The green circles and red squares display high-precision Monte Carlo results of Fig 2 from ref. \cite{Andreas-Frustrated-Ising-Square-1}. Orange and blue triangles display data from the main panel and were added for comparison. As can be observed, the resulting temperatures are in excellent agreement with the reference values despite training only with data from a very limited region of the phase diagram, indicated by thick vertical red lines at $J_2=0$ and $J_2=J$.}
\end{figure*}

On the other hand, we trained a CNN at $J_2 = J$, i.e. far from the maximal frustration point, $J_2/J=0.5$, excluding as before the transition zone.
The transition between the collinear and disordered phases exhibits a different nature (first or second-order transition depending on $J_2$) and has
been studied in detail \cite{Andreas-Frustrated-Ising-Square-1, Andreas-Frustrated-Ising-Square-2,Ising-square-Coll-Sandvik-1}.
Following the same procedure as before, we evaluated the CNN performance by varying $J_2 / J$ from 1 up to $J_2 / J \simeq 0.5$.
Results are shown in the right panel of Fig. \ref{fig:transfer-square-juntas} and a behavior similar to the left panel can be observed but on the other
side of the transition.

With the transition temperatures predicted by the CNN, shown in Fig. \ref{fig:transfer-square-juntas}, we built the temperature vs. frustration phase diagram
of the frustrated Ising model on the square lattice, presented in Fig. \ref{fig:pd-inset-andreas}.
There are three phases, where the system presents N\'eel-order, collinear-order, or is disordered (paramagnetic phase).
The blue (orange) line that separates the N\'eel-ordered (collinear-ordered) phase from the paramagnetic phase is constructed with the critical temperatures
predicted by the CNN trained only in the region indicated by the thick vertical red lines on the left (right), with $J_2=0$ ($J_2=J$).
The CNN-predicted critical temperatures denoted here by blue (orange) triangles correspond to the intersections at $p=0.5$ from the left (right) panel
of Fig. \ref{fig:transfer-square-juntas}.
The inset in Fig. \ref{fig:pd-inset-andreas} shows the same transition temperatures, plotted with the scale of Fig. 2 of ref. \cite{Andreas-Frustrated-Ising-Square-1}, along with that Figure.
The comparison made in the inset shows the excellent agreement between our CNN predictions and the high-precision Monte Carlo results of ref. \cite{Andreas-Frustrated-Ising-Square-1}, indicated with green circles and red squares.
This is remarkable considering that almost the complete phase diagram was obtained by transfer learning since only spin configurations at temperatures within a restricted region (thick vertical red lines) were used as training data.

This result shows the robustness of the CNNs to generalize their predictions to different models from those
in which they were trained, performing equally well in phase transitions of different nature, as illustrated in this example for the
frustrated Ising model on the square lattice.

In the following section we further test the generalization power of CNNs by simultaneously performing a transfer in model and lattice.

\section{Training III: Model-and-lattice transfer}

We conclude the transfer learning study by analyzing the generalization power of a CNN trained with a given lattice and model and evaluating
its performance on another lattice and another model simultaneously, i.e \textit{model-and-lattice transfer}.

Keeping the number of training parameters as low as possible, the network can be forced to learn only the main features from the training data corresponding
to a certain model and lattice. In this way, it is capable of correctly predict the critical temperature from data corresponding to a different model and
lattice (see details in the Appendix).

We train a CNN with Monte Carlo spin configurations of the AFM nearest-neighbor Ising model on the honeycomb lattice excluding the phase transition zone,
and we use it to predict the transition temperatures in the frustrated Ising model on the square lattice.
As we map the honeycomb lattice to a square array (see Appendix \ref{sec:mapping}), the paramagnetic phase in both systems looks the same to the network, but the two low-temperature orderings are different. Hence, in general, a
trained model cannot classify correctly the N\'eel ordered snapshots from the square lattice. Nonetheless, performing several pieces of training it is possible to obtain a model that achieves this task, and the number of tries necessary was found to be no greater than 80, which takes no longer than 2 hours.
To find this model the process is as follows: Once the network is trained on the honeycomb lattice, we use it to predict the order-disorder
probabilities on the square lattice at some fixed value of $J_2$, say $J_2=0.1$. If these two probabilities do not cross each other at any temperature,
the trained model does not predict a transition, it is discarded, and a new training on the honeycomb lattice is performed. If, in contrast,
the two probabilities cross each other, the model is saved. We emphasize here that this selection process does not utilize any a priori knowledge on the
transition temperatures in the square lattice.

\begin{figure}[t!]
\begin{center}
\includegraphics[width=0.9\textwidth]{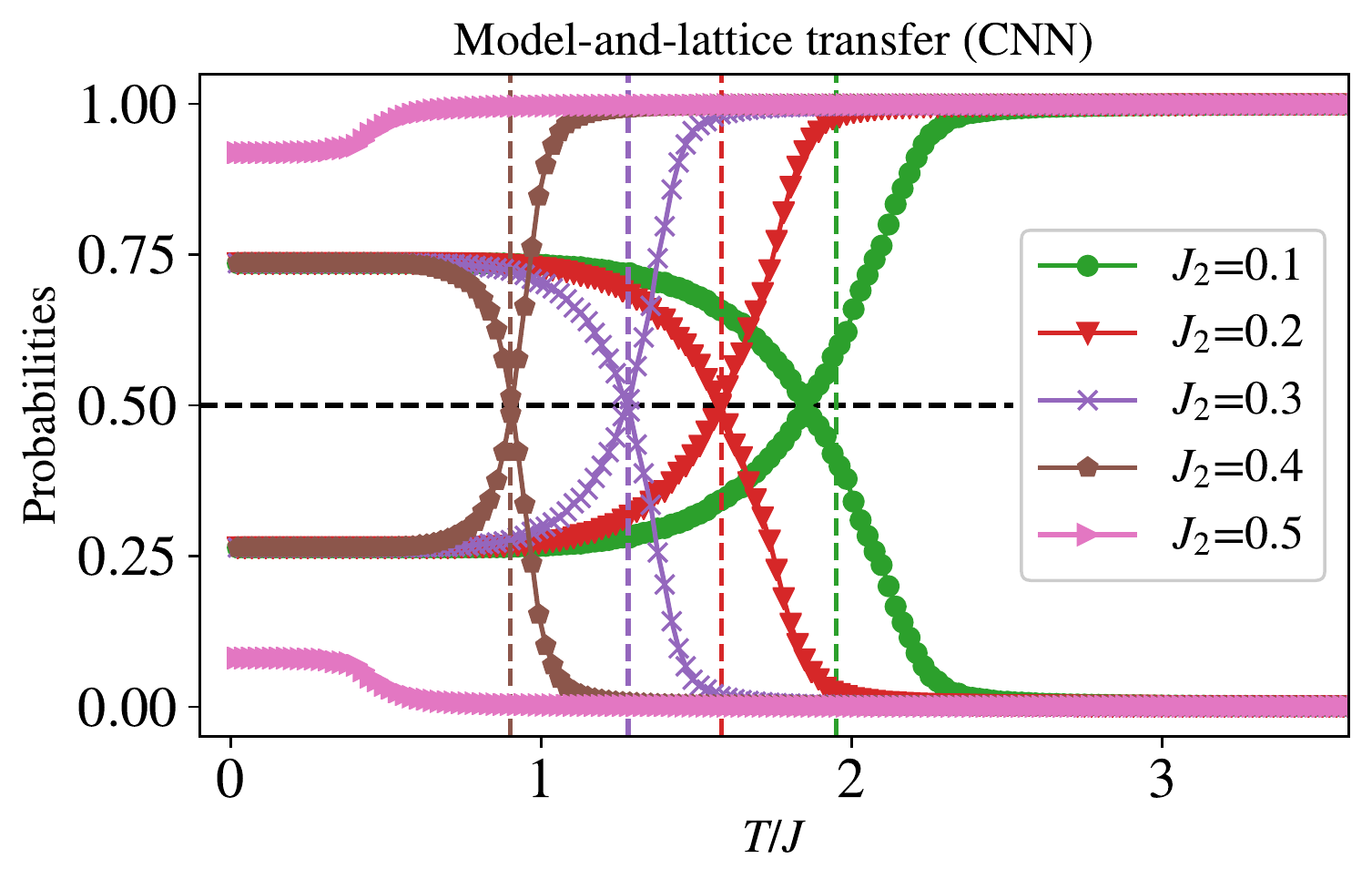}
\end{center}
\caption{\label{fig:latticetransfer1}
Model-and-lattice transfer: CNN output layer probabilities on the \textit{square lattice} as a function of temperature for different frustration values $J_2$. The CNN was only trained with the $J_{2}=0$ \textit{honeycomb lattice} data, with temperatures in the range $0.02 < T/J < 1.22$ and $1.82 < T/J < 4.53$ (the same as in Fig. \ref{fig:AFM-Hon-NN-Ventana}). Vertical dotted lines indicate the corresponding Monte Carlo estimation of transition temperatures. In the legend the $J_2$ values are in units of $J$. Validation accuracy: 0.99.}
\end{figure}

Fig. \ref{fig:latticetransfer1} shows the results of the CNN predictions on the frustrated square lattice for different frustration values $J_2$ as a
function of temperature. As before, the dotted lines indicate the corresponding Monte Carlo estimation of the critical temperatures. The CNN's $T_c$
estimations are very close to those of the Monte Carlo simulations up to $J_2$ values near to the transition between the N\'eel and the collinear phases
at $J_2=0.5 J$ \cite{Andreas-Frustrated-Ising-Square-1}.

%

\section{Training IV: Learning frustration by confusion}
\label{sec:confusion}

To further explore the magnetic phases in frustrated systems via deep learning, in this section we discuss the ``confusion" learning
technique \cite{confusion}.
This method can be considered a hybrid between supervised and unsupervised learning methods.
The central idea of confusion learning is based on evaluating the accuracy, i.e. the ratio of correct predictions over the total number of predictions,
of a network trained with a set of intentionally incorrect labels.
For completeness, we briefly describe the method. Supposing that data depends on a parameter that lies in the range $[a,b]$, in which there is a phase transition at the critical point $c^*$, the method consists of proposing an arbitrary critical point $c$ and training the network by giving the label 0 to all data with parameters smaller than $c$, and giving the label 1 to the rest. Next, the accuracy of the trained network on the complete
training set, $P(c)$, is evaluated with respect to the proposed critical point $c$.
%

\begin{figure}[t!]
\begin{center}
\includegraphics[width=0.9\textwidth]{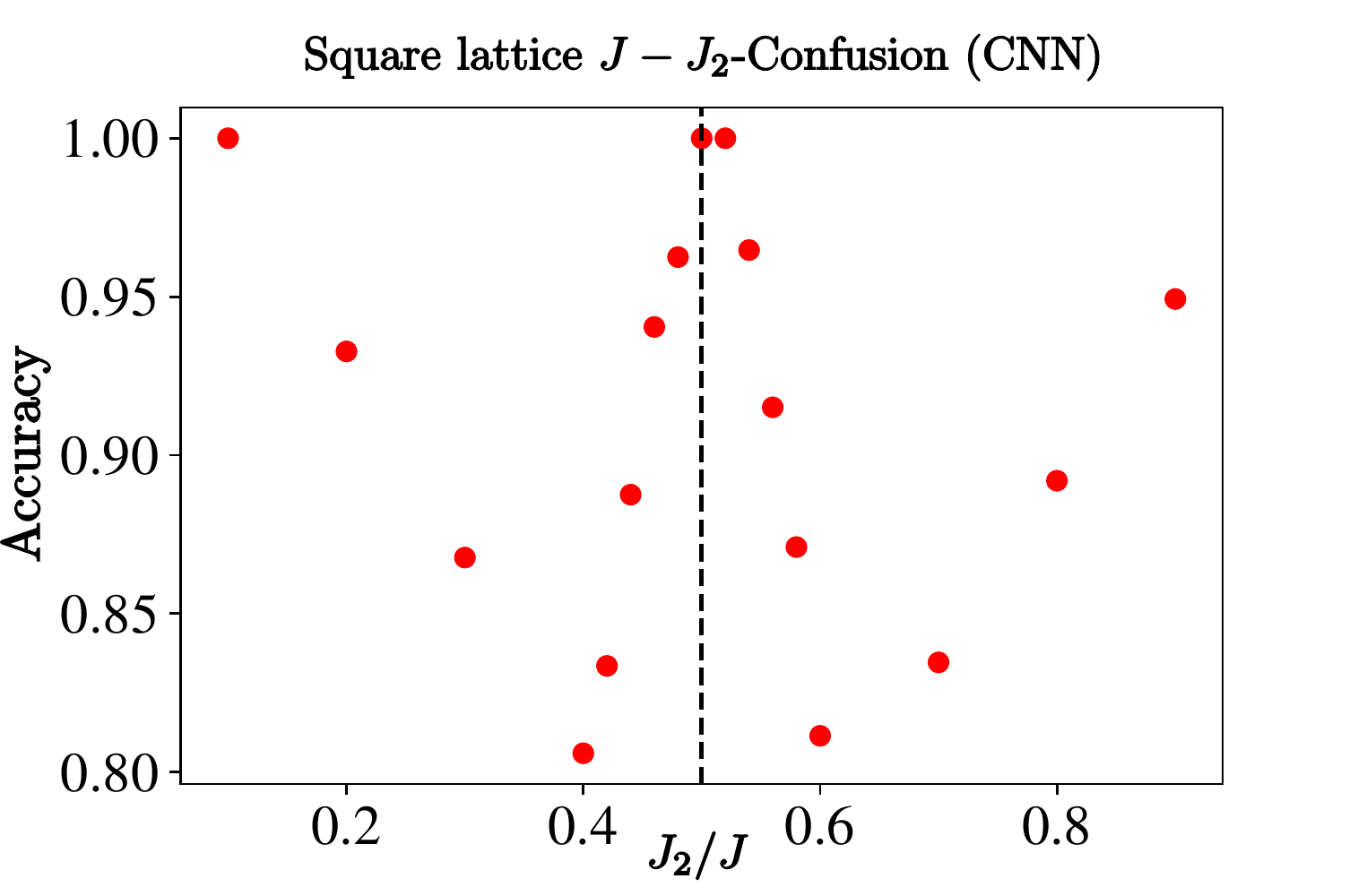}
\end{center}
\caption{\label{fig:confusionSQJ1J2}
Learning frustration by confusion:
Results were obtained by applying the confusion method to the low-temperature phases of the frustrated $J-J_2$ AFM Ising model on the square lattice. The training was implemented at $T/J = 0.02$, indicated by the horizontal red dashed line in Fig. \ref{fig:pd-inset-andreas}. The central peak of the characteristic ``W-shape" is very close to 1/2, which reflects the zero-temperature phase transition at $J_2/J=1/2$ from the N\'eel to the collinear phase. This indicates that the confusion method is a powerful technique to determine phase transitions even in the presence of frustration.}
\end{figure}

By sweeping the $c$ values and repeating the process over the range $[a,b]$, the function $P(c)$ is obtained, which exhibits a W-shape with its intermediate
peak located at the true critical point $c^*$.  \\
The reason for this is as follows. When $c = a$ or $c=b$ all data is labeled with the same label, and therefore the network predicts with $100\%$ accuracy.
This explains the two ends of the W-shape.
The central maximum in the network performance occurs when the proposed critical point $c$ coincides with the true critical point $ c^* $.
In this case, the training process is equivalent to the standard supervised learning of the network.

Fig. \ref{fig:confusionSQJ1J2} shows the results obtained by applying the confusion method to the low-temperature phases of the frustrated
$J-J_2$ AFM Ising model on the square lattice. In this case, the training process was performed at $T/J = 0.02$, indicated by the horizontal
red dashed line in Fig. \ref{fig:pd-inset-andreas} (see the Appendix for the CNN implementation details).
For $T \rightarrow 0$, the model presents a phase transition at $J_2/J=1/2$ from N\'eel to collinear order \cite{Andreas-Frustrated-Ising-Square-1}.
Note how the central peak of the W-shape emerges very close to 1/2 in Fig. \ref{fig:confusionSQJ1J2}, indicating that the method can precisely determine
the transition.

It is important to recall that the confusion method can detect changes in data patterns that are not necessarily associated with a physical phase transition.
In this sense, the method is an indicator of a transition, which should be contrasted with other methods or prior knowledge of the physics involved.

%

\section{Conclusions}
\label{sec:conclusions}

In this work, we have explored the ability of neural networks to generalize knowledge beyond their training. This aspect of transfer learning,
which has become a cornerstone in data science, has been discussed here in the context of antiferromagnetic Ising spin systems.

The transfer learning analysis was carried out in three stages of increasing order of generality demand.

We call the first case \textit{local transfer}. Here, given a model on a lattice, we train a neural network to distinguish spin configurations far from the corresponding order-disorder transition temperature.
Then we analyze the performance of the neural network to differentiate both phases in the complete range of temperatures, and we determine the transition temperature. This target is outside the training zone, although within the same model and lattice, hence the term ``local" transfer.
Our result is as follows. For non-frustrated lattices,  \textit{local transfer} works properly with DNNs. In this work, we illustrate this case with the AFM first neighbors Ising model on the honeycomb lattice (Fig. \ref{fig:AFM-Hon-NN-Ventana}). However, for frustrated lattices, DNNs are not sufficiently accurate to guarantee adequate
local transfer. This was tested not only with spin configurations but also with correlations as input features to train the neural network. We exemplified this case with the AFM first neighbors Ising model on the triangular lattice. The structure of the lattice gives rise in this case to frustration and consequently high degeneracy. This makes it difficult for the DNN to classify properly (Fig. \ref{fig:AFM-triang-NN-Ventana}). This difficulty is overcome using a convolutional network, whose pre-processing filter architecture extracts more representative features directly from the image of the configurations, making it suitable in high degeneracy cases (Fig. \ref{fig:AFM-traing-CNN-Ventana}).

The second case analyzed was \textit{model transfer}, in which we train a CNN using data on a given lattice (excluding the transition zone) and we then test its capability to classify data of a more general model on the same lattice. We illustrated this case with three examples. In the first one, we train the CNN with data corresponding to the non-frustrated AFM Ising model on the honeycomb lattice and we use it to classify data adding finite next-neighbors antiferromagnetic interactions

(Fig. \ref{fig:model_tranfer_honeycombe_J2_Neel}). For the other two examples we carried out a similar procedure, but for the AFM Ising model on the square lattice. We train the CNN
far away from the maximally frustrated point $J_2=J/2$, i.e., with data with $J_2=0$, or $J_2=J$, respectively.
With these two pieces of training, we construct the phase diagram of Fig. \ref{fig:pd-inset-andreas}. The agreement of our CNN results with high-precision Monte Carlo results \cite{Andreas-Frustrated-Ising-Square-1} depicted in the figure inset is significant, considering the small region of training.

These results indicate that CNNs generalize appropriately from the features learned in the restricted models where they were trained. The CNN not only quantitatively identifies the transitions between the ordered and high-temperature phases but also gives evidence of the transitions between the low-temperature phases.

Finally, for the third and most demanding case, we considered \textit{model-and-lattice transfer}.
To this end, we train a convolutional network with spin configurations corresponding to the non-frustrated AFM Ising model on the honeycomb lattice, and we
test the CNN performance on the frustrated AFM Ising model on the square lattice.
The results in Fig. \ref{fig:latticetransfer1} show that it is possible to find a model that can generalize adequately,
finding the order-disorder transitions for different frustration values, and signaling the low-temperature phase transition induced by frustration.

The transfer learning results presented in this work indicate that neural networks, in particular convolutional networks, can be adequate generic classifiers, exhibiting high performance when properly trained in minimal architectures, even in cases of high degeneracy such as the frustrated systems already analyzed. We plan to apply similar methods to other frustrated models at classical \cite{Arlego-Honecker-Kalz} and quantum level \cite{Lamas-1,Lamas-3,Lamas-Arlego-Zhang-Brenig}.

In addition to the implementation of supervised transfer learning, in this paper, we addressed the  ``learning by confusion" technique.

We implemented learning by confusion on the $J-J_2$ model on the square lattice using a CNN. Our results, depicted in Fig. \ref{fig:confusionSQJ1J2}, clearly show that the method can detect the emergence of the transition between the low-temperature phases. This is evidenced by the intermediate peak of the characteristic ``W" shape of the predicted accuracy in Fig. \ref{fig:confusionSQJ1J2}, which is located very close to the transition point at $J_2 = 0.5 J$.

The previous example highlights the main advantage of the confusion method, i.e. not having to rely on correct labeling beforehand to detect pattern
changes in the configuration data. In this case, the neural network learns to differentiate the low-temperature phases when frustration generates high
degeneracy. In this sense, learning by confusion offers a complementary tool to the supervised learning techniques presented above.

\section*{Acknowledgments}
This research was partially supported by CONICET (P-UE 22920170100066CO), UNLP (Grant No. 11/X678, 11/X806).
We would like to thank Andreas Honecker and Mateus Schmidt for very valuable information exchange.
We also thank Andreas Honecker, Ansgar Kalz, and Marion Moliner for allowing us to use one of their Figures.


\bibliographystyle{elsarticle-num}
\bibliography{refs-2020}

\begin{thebibliography}{10}
\expandafter\ifx\csname url\endcsname\relax
  \def\url#1{\texttt{#1}}\fi
\expandafter\ifx\csname urlprefix\endcsname\relax\def\urlprefix{URL }\fi
\expandafter\ifx\csname href\endcsname\relax
  \def\href#1#2{#2} \def\path#1{#1}\fi

\bibitem{lecun2015deep}
Y.~LeCun, Y.~Bengio, G.~Hinton, Deep learning, nature 521~(7553) (2015)
  436--444.

\bibitem{DeepLearningBible}
I.~Goodfellow, Y.~Bengio, A.~Courville, Deep learning, MIT press, 2016.

\bibitem{hastie2009elements}
T.~Hastie, R.~Tibshirani, J.~Friedman, The elements of statistical learning,
  Springer Science \& Business Media, 2009.

\bibitem{language-power-law-corr}
W.~Li, Random texts exhibit zipf's-law-like word frequency distribution, IEEE
  Transactions on information theory 38~(6) (1992) 1842--1845.

\bibitem{Phase-Transitions-in-ML-Book}
L.~Saitta, A.~Giordana, A.~Cornu{\'e}jols,
  \href{https://books.google.com.ar/books?id=wG4UnmfsdWQC}{Phase Transitions in
  Machine Learning}, Cambridge University Press, 2011.
\newline\urlprefix\url{https://books.google.com.ar/books?id=wG4UnmfsdWQC}

\bibitem{MEHTA20191}
P.~Mehta, M.~Bukov, C.-H. Wang, A.~G. Day, C.~Richardson, C.~K. Fisher, D.~J.
  Schwab,
  \href{http://www.sciencedirect.com/science/article/pii/S0370157319300766}{A
  high-bias, low-variance introduction to machine learning for physicists},
  Physics Reports 810 (2019) 1 -- 124.
\newblock \href {https://doi.org/https://doi.org/10.1016/j.physrep.2019.03.001}
  {\path{doi:https://doi.org/10.1016/j.physrep.2019.03.001}}.
\newline\urlprefix\url{http://www.sciencedirect.com/science/article/pii/S0370157319300766}

\bibitem{carleo2019machine}
G.~Carleo, I.~Cirac, K.~Cranmer, L.~Daudet, M.~Schuld, N.~Tishby,
  L.~Vogt-Maranto, L.~Zdeborov{\'a}, Machine learning and the physical
  sciences, Reviews of Modern Physics 91~(4) (2019) 045002.

\bibitem{Carrasquilla-Review}
J.~{Carrasquilla}, {Machine Learning for Quantum Matter}, arXiv e-prints (2020)
  arXiv:2003.11040\href {http://arxiv.org/abs/2003.11040}
  {\path{arXiv:2003.11040}}.

\bibitem{carrasquilla2017machine}
J.~Carrasquilla, R.~G. Melko, Machine learning phases of matter, Nature Physics
  13~(5) (2017) 431--434.

\bibitem{wang2017machine}
C.~Wang, H.~Zhai, Machine learning of frustrated classical spin models. i.
  principal component analysis, Physical Review B 96~(14) (2017) 144432.

\bibitem{wang2018machine}
C.~Wang, H.~Zhai, Machine learning of frustrated classical spin models (ii):
  Kernel principal component analysis, Frontiers of Physics 13~(5) (2018) 1--7.

\bibitem{beach2018machine}
M.~J. Beach, A.~Golubeva, R.~G. Melko, Machine learning vortices at the
  kosterlitz-thouless transition, Physical Review B 97~(4) (2018) 045207.

\bibitem{ch2017machine}
K.~Ch’Ng, J.~Carrasquilla, R.~G. Melko, E.~Khatami, Machine learning phases
  of strongly correlated fermions, Physical Review X 7~(3) (2017) 031038.

\bibitem{zhang2017machine}
Y.~Zhang, R.~G. Melko, E.-A. Kim, Machine learning z 2 quantum spin liquids
  with quasiparticle statistics, Physical Review B 96~(24) (2017) 245119.

\bibitem{Potts2020}
K.~Shiina, H.~Mori, Y.~Okabe, H.~K. Lee, Machine-learning studies on spin
  models, Scientific reports 10~(1) (2020) 1--6.

\bibitem{ponte2017kernel}
P.~Ponte, R.~G. Melko, Kernel methods for interpretable machine learning of
  order parameters, Physical Review B 96~(20) (2017) 205146.

\bibitem{suchsland2018parameter}
P.~Suchsland, S.~Wessel, Parameter diagnostics of phases and phase transition
  learning by neural networks, Physical Review B 97~(17) (2018) 174435.

\bibitem{greitemann2019probing}
J.~Greitemann, K.~Liu, L.~Pollet, et~al., Probing hidden spin order with
  interpretable machine learning, Physical Review B 99~(6) (2019) 060404.

\bibitem{ch2018unsupervised}
K.~Ch'ng, N.~Vazquez, E.~Khatami, Unsupervised machine learning account of
  magnetic transitions in the hubbard model, Physical Review E 97~(1) (2018)
  013306.

\bibitem{broecker2017quantum}
P.~Broecker, F.~F. Assaad, S.~Trebst, Quantum phase recognition via
  unsupervised machine learning, arXiv preprint arXiv:1707.00663 (2017).

\bibitem{huembeli2018identifying}
P.~Huembeli, A.~Dauphin, P.~Wittek, Identifying quantum phase transitions with
  adversarial neural networks, Physical Review B 97~(13) (2018) 134109.

\bibitem{liu2019learning}
K.~Liu, J.~Greitemann, L.~Pollet, et~al., Learning multiple order parameters
  with interpretable machines, Physical Review B 99~(10) (2019) 104410.

\bibitem{hsu2018machine}
Y.-T. Hsu, X.~Li, D.-L. Deng, S.~D. Sarma, Machine learning many-body
  localization: Search for the elusive nonergodic metal, Physical Review
  Letters 121~(24) (2018) 245701.

\bibitem{broecker2017machine}
P.~Broecker, J.~Carrasquilla, R.~G. Melko, S.~Trebst, Machine learning quantum
  phases of matter beyond the fermion sign problem, Scientific reports 7~(1)
  (2017) 1--10.

\bibitem{zhang2019interpretable}
W.~Zhang, L.~Wang, Z.~Wang, Interpretable machine learning study of the
  many-body localization transition in disordered quantum ising spin chains,
  Physical Review B 99~(5) (2019) 054208.

\bibitem{ni2019machine}
Q.~Ni, M.~Tang, Y.~Liu, Y.-C. Lai, Machine learning dynamical phase transitions
  in complex networks, Physical Review E 100~(5) (2019) 052312.

\bibitem{rem2019identifying}
B.~S. Rem, N.~K{\"a}ming, M.~Tarnowski, L.~Asteria, N.~Fl{\"a}schner,
  C.~Becker, K.~Sengstock, C.~Weitenberg, Identifying quantum phase transitions
  using artificial neural networks on experimental data, Nature Physics 15~(9)
  (2019) 917--920.

\bibitem{wetzel2020discovering}
S.~J. Wetzel, R.~G. Melko, J.~Scott, M.~Panju, V.~Ganesh, Discovering symmetry
  invariants and conserved quantities by interpreting siamese neural networks,
  arXiv preprint arXiv:2003.04299 (2020).

\bibitem{wang2016discovering}
L.~Wang, Discovering phase transitions with unsupervised learning, Physical
  Review B 94~(19) (2016) 195105.

\bibitem{Auerbach:1994}
A.~Auerbach, Interacting Electrons and Quantum Magnetism, Graduate Texts in
  Contemporary Physics, Springer-Verlag New York, 1994.
\newblock \href {https://doi.org/10.1007/978-1-4612-0869-3}
  {\path{doi:10.1007/978-1-4612-0869-3}}.

\bibitem{Book-Computational-Many-Particle-Physics}
H.~Fehske, R.~Schneider, A.~Wei{\ss}e,
  \href{https://books.google.com.ar/books?id=2TugcQAACAAJ}{Computational
  Many-Particle Physics}, Lecture Notes in Physics, Springer Berlin Heidelberg,
  2010.
\newline\urlprefix\url{https://books.google.com.ar/books?id=2TugcQAACAAJ}

\bibitem{chollet2017-Book}
F.~Chollet, \href{https://books.google.com.ar/books?id=Yo3CAQAACAAJ}{Deep
  Learning with Python}, Manning Publications Company, 2017.
\newline\urlprefix\url{https://books.google.com.ar/books?id=Yo3CAQAACAAJ}

\bibitem{Landau-Binder-MC-Book}
D.~P. Landau, K.~Binder, A guide to Monte Carlo simulations in statistical
  physics, Cambridge university press, 2014.

\bibitem{chollet2015keras}
F.~Chollet, et~al., Keras, \url{https://github.com/fchollet/keras} (2015).

\bibitem{confusion}
E.~P. Van~Nieuwenburg, Y.-H. Liu, S.~D. Huber, Learning phase transitions by
  confusion, Nature Physics 13~(5) (2017) 435--439.

\bibitem{onsager1944crystal}
L.~Onsager, Crystal statistics. i. a two-dimensional model with an
  order-disorder transition, Physical Review 65~(3-4) (1944) 117.

\bibitem{Ising-Exact-Honeycomb}
R.~M.~F. Houtappel, Order-disorder in hexagonal lattices, Physica 16~(5) (1950)
  425--455.

\bibitem{wannier1950-Ising-AFM-Triangular}
G.~Wannier, Antiferromagnetism. the triangular ising net, Physical Review
  79~(2) (1950) 357.

\bibitem{Frustrated-Magnetism-book}
C.~Lacroix, P.~Mendels, F.~Mila, Introduction to frustrated magnetism, Vol.
  164, Springer Science \& Business Media, 2011.

\bibitem{Ising-Honeycomb-frustrated}
A.~Bob{\'a}k, T.~Lu{\v{c}}ivjansk{\`y}, M.~{\v{Z}}ukovi{\v{c}},
  M.~Borovsk{\`y}, T.~Balcerzak, Tricritical behaviour of the frustrated ising
  antiferromagnet on the honeycomb lattice, Physics Letters A 380~(34) (2016)
  2693--2697.

\bibitem{Andreas-Frustrated-Ising-Square-1}
A.~Kalz, A.~Honecker, M.~Moliner, Analysis of the phase transition for the
  ising model on the frustrated square lattice, Physical Review B 84~(17)
  (2011) 174407.

\bibitem{Andreas-Frustrated-Ising-Square-2}
A.~Kalz, A.~Honecker, Location of the potts-critical end point in the
  frustrated ising model on the square lattice, Physical Review B 86~(13)
  (2012) 134410.

\bibitem{Ising-square-Coll-Sandvik-1}
S.~Jin, A.~Sen, A.~W. Sandvik, Ashkin-teller criticality and pseudo-first-order
  behavior in a frustrated ising model on the square lattice, Physical Review
  Letters 108~(4) (2012) 045702.

\bibitem{Arlego-Honecker-Kalz}
A.~Kalz, M.~Arlego, D.~Cabra, A.~Honecker, G.~Rossini,
  \href{https://link.aps.org/doi/10.1103/PhysRevB.85.104505}{Anisotropic
  frustrated heisenberg model on the honeycomb lattice}, Phys. Rev. B 85 (2012)
  104505.
\newblock \href {https://doi.org/10.1103/PhysRevB.85.104505}
  {\path{doi:10.1103/PhysRevB.85.104505}}.
\newline\urlprefix\url{https://link.aps.org/doi/10.1103/PhysRevB.85.104505}

\bibitem{Lamas-1}
D.~C. Cabra, C.~A. Lamas, H.~D. Rosales,
  \href{https://link.aps.org/doi/10.1103/PhysRevB.83.094506}{Quantum disordered
  phase on the frustrated honeycomb lattice}, Phys. Rev. B 83 (2011) 094506.
\newblock \href {https://doi.org/10.1103/PhysRevB.83.094506}
  {\path{doi:10.1103/PhysRevB.83.094506}}.
\newline\urlprefix\url{https://link.aps.org/doi/10.1103/PhysRevB.83.094506}

\bibitem{Lamas-3}
C.~A. Lamas, A.~Ralko, M.~Oshikawa, D.~Poilblanc, P.~Pujol,
  \href{https://link.aps.org/doi/10.1103/PhysRevB.87.104512}{Hole statistics
  and superfluid phases in quantum dimer models}, Phys. Rev. B 87 (2013)
  104512.
\newblock \href {https://doi.org/10.1103/PhysRevB.87.104512}
  {\path{doi:10.1103/PhysRevB.87.104512}}.
\newline\urlprefix\url{https://link.aps.org/doi/10.1103/PhysRevB.87.104512}

\bibitem{Lamas-Arlego-Zhang-Brenig}
H.~Zhang, C.~A. Lamas, M.~Arlego, W.~Brenig,
  \href{https://link.aps.org/doi/10.1103/PhysRevB.97.235123}{Nematic quantum
  phases in the bilayer honeycomb antiferromagnet}, Phys. Rev. B 97 (2018)
  235123.
\newblock \href {https://doi.org/10.1103/PhysRevB.97.235123}
  {\path{doi:10.1103/PhysRevB.97.235123}}.
\newline\urlprefix\url{https://link.aps.org/doi/10.1103/PhysRevB.97.235123}

\end{thebibliography}

\appendix

\section{Methods}

The computations performed in this paper are open source and written in C and Python, using specific libraries such as Keras for deep learning.

\noindent {\it Monte Carlo simulations:} The Monte Carlo generation of tagged data, using a Metropolis Algorithm and single-spin-flip dynamics, is performed as follows.
We run 400 independent simulations starting from the high-temperature phase (in general $T_0=4.5 J$). A set of 200 evenly-spaced temperature values is obtained from the range $[0,T_0]$, and for each temperature, the spin configuration and the temperature are saved once equilibrium is reached. Thus, our dataset for each lattice and each model considered consists of 80000 samples.

We use the analytical expressions available for the critical temperature in the cases of the non-frustrated Ising model on the square and honeycomb lattices.
In the other cases, the location of the Monte Carlo specific heat maximum, for $N=900$ sites, is taken as an estimator for the order-disorder transitions.
In this work, this procedure was carried out for the frustrated Ising model on the square, honeycomb, and triangular lattices.
In the latter, the above-mentioned maximum does not represent an order-disorder transition. In this case, we identify this temperature as $T^*$ to avoid confusion.

\noindent {\it Training procedure:} The data generated by Monte Carlo simulations is split 70 \% for training (10 \%  of which is taken
for validation) and 30 \% for the test, i.e. prediction. Data with $T < T_c $ is labeled with 0 and data with $T > T_c $ is labeled with 1. The same exclusion process is carried out around $T^*$ for the triangular lattice.
Temperature values are only used in the test stage, to analyze the performance of the classification, and to estimate the critical temperature.
We have trained the network in a range of temperatures that excludes a window $|T-T_c|<w J$, with $ 0.1\leq w \leq 0.3$.
Therefore, no information about the critical region is introduced during training (\textit{local transfer}).
For \textit{model transfer} and \textit{model-and-lattice transfer} we used a similar procedure but validating on a different model as well as a different model and lattice to the training one, respectively. In the range of windows analyzed there are no significant variations in the accuracy.

\noindent {\it Dense Neural Networks (DNN):} In Fig. 2 we use as input a vector with the local spin configurations that we compute from the Monte Carlo simulations, normalized to 0 (spin down) or 1 (spin up). Three DNNs have an input layer having $N = 100, 400$ and $900$ nodes, corresponding to three system sizes,
respectively. The single hidden layer contains 16 neurons with ReLu activation functions.  The stochastic optimization method is Adam, and the loss function is categorical cross-entropy. For training, we use roughly 50000-70000 configurations, 10 \% of which is used for validation during training.
The learning rate is order $10^{-4}$, the batch size is 128, the $L_2$ regularization factor is order $10^{-5}$, and the number of epochs is 10, giving a final validation
accuracy of 0.99 and a validation loss of order $10^{-2}$. The output layer has two neurons (binary classification), with Softmax activation functions.
The probability curves in Figs. 2 and 3 are the result of the average over roughly 150 independent samples from the test set for each temperature.
In Fig. 3 both spin configurations and spin correlations input variables are normalized to 0 or 1. The hyperparameters that changed with respect to the
previously mentioned are the number of neurons in the hidden layer, increased from 16 to 32, and the number of epochs increased from 10 to 20.
The final validation accuracy is roughly 0.98 and there is a validation loss of order $10^{-1}$ over the training dataset. 	

\noindent {\it Convolutional Neural Networks (CNN):} The input is a 30x30 matrix, with normalized spin configurations. Data segmentation in train, validation, and
test sets, is made as in the DNN case. Several architectures are suitable for classification tasks. The CNNs used consist of firstly in one or two
convolutional layers of 3 to 10 filters of size 3x3, each followed by a max- or average-pooling layer. Then, the data is flattened to a one-dimensional
vector which is the input of a dense layer with 3 to 16 neurons with ReLu activation functions. The optimization method is Adam, and the loss function is again
the categorical cross-entropy. The batch sizes are between 128 and 512, the number of epochs is less than 5 to prevent overfitting, and the learning rate
is order $10^{-3}-10^{-4}$. The output layer has again two neurons with Softmax activation functions. Validation accuracy in training is higher than 0.99 and validation loss is order $10^{-2}$. As before, the probability predictions over test datasets are averaged over roughly 150 independent samples from the test set for each temperature.

Source data for all the figures in the paper and all training and test data used are available upon request from the authors.
%
Source code for training and evaluating our neural networks is available upon request from the authors.
%

\section{Mapping the Honeycomb lattice  to a square array}
\label{sec:mapping}

\begin{figure}[t!]
\begin{center}
 \includegraphics[width=.8\linewidth]{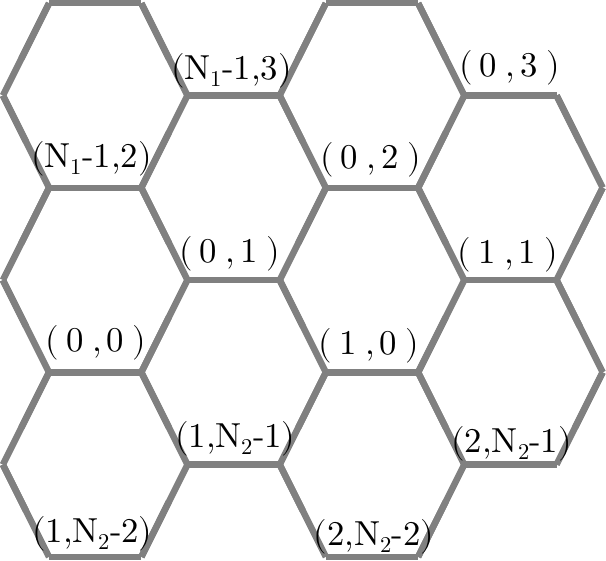}
\end{center}
\caption{Honeycomb lattice. Ordered pairs correspond to the indexation for each cell used in this work. }
\label{fig:honeycomb}
\end{figure}

The size of the system is 900 sites and the Monte Carlo simulation was performed using periodic boundary conditions.
Each unit cell is indexed by two integers $(i,j)$, where $0\leq i<N_1=30$ and $0 \leq j<N_2=15$ as we show in Fig. \ref{fig:honeycomb}.

We map the honeycomb lattice to a square $30 \times 30 $ array $A$  as follows:
Each spin in the honeycomb lattice is indexed by three integers in the 3-rank tensor $S_{ij}^k$, where $i$ and $j$ determine the unit cell $(i,j)$,
and $k=0$ or $k=1$ corresponds to the left or right spin in the unit cell, respectively.
Then, we can construct the square $30 \times 30 $ array $A$ by taking

\begin{equation}
 A_{mn}=S_{m \floor*{n/2}}^{mod(n,2)},
 \label{eq:map}
\end{equation}
with $0\leq m,n<30$, being $mod(n,2)$ the rest in the division of $n$ by two, and $\floor*{n/2}$ the floor function which gets the integer part
in the division.

\begin{figure}[t!]
\begin{center}
 \includegraphics[width=.7\linewidth]{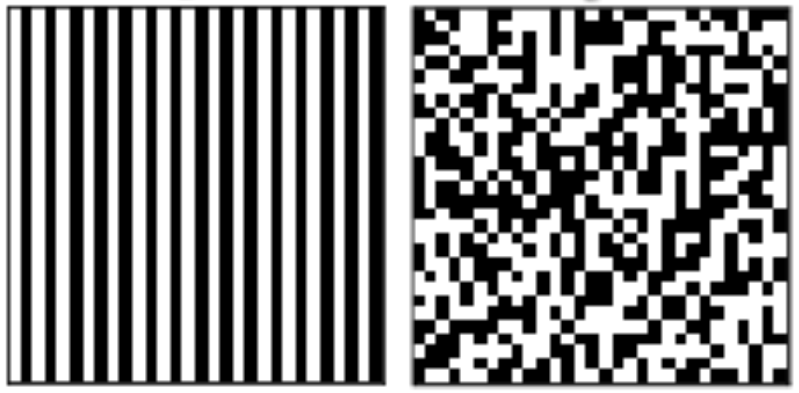}
\end{center}
\caption{Snapshots of the system mapped to a $30 \times 30 $ pixel image. Each black or white pixel corresponds to a down or up spin, respectively. The left panel corresponds to a N\'eel ordered spin configuration at $T=0.02$ whereas the right panel corresponds to a disordered spin configuration at $T=4.53$. }
\label{fig:snapshotss}
\end{figure}

Using the mapping \eqref{eq:map}, we can visualize snapshots of the system in a 2D image as we show in Fig \ref{fig:snapshotss}.
The left panel corresponds to a N\'eel ordered spin configuration at $T=0.02$ whereas the right panel corresponds to a disordered spin
configuration at $T=4.53$.

\end{document}